\documentclass[journal, twoside, web]{ieeecolor}
\usepackage{lcsys}

\usepackage[utf8]{inputenc}
\usepackage[T1]{fontenc}
\usepackage{amsmath}
\usepackage{amsfonts}
\usepackage{amssymb}
\usepackage{graphicx}
\usepackage{physics}
\usepackage{commath}
\usepackage{xcolor}
\usepackage{float}
\usepackage{stfloats}
\usepackage{optidef}
\usepackage{subfig}
\usepackage{mathtools}
\DeclarePairedDelimiterX{\inp}[2]{\langle}{\rangle}{#1, #2}

\usepackage{tikz}
\usepackage{pgfplots}
\usepgfplotslibrary{groupplots} 
\usepgfplotslibrary{colorbrewer}
\usetikzlibrary{calc}

\pgfkeys{/pgf/number format/.cd,1000 sep={}} 
\pgfplotsset{every axis/.append style={yticklabel style={/pgf/number format/fixed,/pgf/number format/precision=5},scaled y ticks=false,}} 
\pgfplotsset{every axis/.append style={xticklabel style={/pgf/number format/fixed,/pgf/number format/precision=5},scaled x ticks=false,}} 

\pgfplotsset{
	wide1/.append style={width=6.5cm, height=3cm,}
}

\pgfplotsset{
	wide2/.append style={width=6.5cm, height=1.5cm,}
}

\pgfplotsset{
	small1/.append style={width=3cm, height=2cm,}
}

\pgfplotsset{
	wide3/.append style={width=7.15cm, height=3cm,}
}

\pgfplotsset{default/.style={wide1,
		scale only axis, 
		cycle list/Dark2,
		ylabel near ticks, xlabel near ticks, xmajorgrids, ymajorgrids,
		enlargelimits=false, enlarge y limits=upper, 
		legend pos=north east, 
		every axis legend/.append style={legend cell align=left,align=left, font=\small, fill opacity=0.7,draw opacity=1,text opacity=1},
		every axis label/.append style={font=\small},
		ticklabel style={font=\small},
		every axis plot/.append style={thick, mark=none},
		every axis title/.append style={font=\bfseries},
	}
}

\usetikzlibrary{external}
\usepackage{cuted} 

\newcommand{\lkhood}{l}

\newcounter{mytempeqncnt}

\newcommand{\revised}[1]{{\color{blue!70!black}#1}}
\renewcommand{\revised}[1]{#1}

\newcommand{\removed}[1]{}

\makeatletter
\let\NAT@parse\undefined
\makeatother
\usepackage[hidelinks]{hyperref}

\begin{document}
\title{Online Parameter Estimation in Continuously Monitored Quantum Systems}
\author{Henrik Glavind Clausen, Pierre Rouchon, Rafal Wisniewski
	\thanks{The work of H. G. Clausen and R. Wisniewski was supported by Independent Research Fund Denmark (DFF), project number 0136-00204B. The work of P. Rouchon was supported by the European Research Council (ERC) under the European Union's Horizon 2020 research and innovation program (grant agreement No. 884762).}%
	\thanks{H. G. Clausen and R. Wisniewski are with Section of Automation and Control, Department of Electronic Systems, Aalborg University, DK-9220 Aalborg Øst, Denmark (email: {\tt \{hgcl,raf\}@es.aau.dk}).}%
	\thanks{P. Rouchon is with Laboratoire de Physique de l'Ecole Normale Supérieure, Mines Paris-PSL, Inria, ENS-PSL, Université PSL, CNRS, Paris, France (email: {\tt pierre.rouchon@minesparis.esl.eu}).}%
}

\maketitle

\thispagestyle{empty}
\pagestyle{empty}

\begin{abstract}
	In this work, we consider the problem of online (real-time, single-shot) estimation of static or slow-varying parameters along quantum trajectories in quantum dynamical systems. Based on the measurement signal of a continuously monitored quantum system, we propose a recursive algorithm for computing the maximum likelihood (ML) estimate of unknown parameters using an approach based on stochastic gradient ascent on the log-likelihood function. We formulate the algorithm in both discrete-time and continuous-time and illustrate the performance of the algorithm through simulations of a simple two-level system undergoing homodyne measurement from which we are able to track multiple parameters simultaneously.
\end{abstract}


\section{Introduction}
\IEEEPARstart{W}{ith} the rapid development of engineered quantum systems, the need for estimation and monitoring of system parameters in quantum dynamical systems is an important task, e.g., when calibrating quantum experiments or for achieving high-precision sensing in quantum metrology.


In this work, we focus on the particular setting when a quantum system is subject to continuous measurements, in which case the resulting process is described by a stochastic master equation (SME), either in discrete-time via a Kraus map formulation or in continuous-time as a diffusion process \cite{wiseman2010,rouchon2022}. \revised{Nowadays, such systems appear routinely in many experimental quantum platforms such as circuit quantum electrodynamics (circuit QED) \cite{blais2021}, atomic magnetometry \cite{amoros-binefa2021}, quantum optics and optomechanics \cite{rossi2019}.} 

In the context of parameter estimation, we distinguish between two paradigms: (i) Offline or batch estimation, where the complete set of data is processed simultaneously to provide an estimate, and (ii) online or recursive estimation, where the estimate is continuously updated as new data arrives. 

Among the primary motivations for studying online algorithms is the ability to track parameter changes as they happen in real-time, but online algorithms may also be useful alternatives to offline processing for very long time-series data for which the computational burden of typical offline algorithms become too heavy.

In general, the optimal parameter estimate, in the Bayesian sense, is given in terms of the solution to the (nonlinear) filtering problem (i.e., a solution to the Kushner-Stratonovich equation \cite{jazwinski}). In the special case where the quantum system at hand can be represented as a linear Gaussian process, the (online) parameter estimation problem can be solved by the Kalman filter, as was illustrated, e.g., for a quantum harmonic oscillator subject to an unknown force \cite{verstraete2001} and for estimating a magnetic field in an atomic spin ensemble \cite{geremia2003,stockton2004,amoros-binefa2021}.

Except for the linear case, the filtering problem rarely has a tractable solution, but it may be approximated in various ways. For offline processing, this has been investigated via discretizing the parameter space \cite{gambetta2001,warszawski2004} and using Markov chain Monte Carlo methods \cite{gammelmark2013,kiilerich2016}, whereas methods leaning towards online processing have been developed by utilizing the quantum structure to embed the discretized parameter space in an enlarged Hilbert space, resulting in the so-called quantum particle filter \cite{chase2009,negretti2013,six2015,ralph2017,bompais2022}. Common to these methods is, however, that they require propagation of a separate quantum filter for each parameter value in a discretized parameter space, resulting in very computationally demanding algorithms, especially for multi-parameter problems. 

A slightly different approach is taken by \cite{ralph2011}, where classical frequency estimation techniques from the signal processing literature are employed, and in \cite{cortez2017}, where an algorithm is proposed for computing the maximum likelihood (ML) estimate without having to propagate a quantum filter.

In this work, we propose an online algorithm for computing the ML parameter estimate with the aim of tracking slowly varying parameters. The method presented in this work relies on a recursive formulation of the gradient of the log-likelihood based on propagation of the quantum filter and the associated sensitivity equations (sometimes referred to as the filter derivative or the tangent filter). A different but commonly used method for computing the gradient of the log-likelihood is based on the smoothing distribution, typically via a forward-backward approach (see, e.g., \cite{six2016}), which does not lend itself straightforwardly to a recursive formulation. By contrast, the approach based on sensitivity equations only requires integration forward in time, making it inherently recursive. For a detailed discussion on the relationship between the two approaches in the context of general hidden Markov models (HMMs), see \cite[Chapter~10]{cappe2005}. 

While this approach is by no means new in the context of classical systems theory (see, e.g., \cite{benveniste1990,ljung1983}), it has received limited attention over the years. Among the latest contributions are \cite{surace2019}, where the authors present and prove convergence of an online ML estimator for general nonlinear diffusion processes subject to continuous-time measurements. As highlighted in \cite{surace2019}, such methods lend themselves best to systems that admit an exact, recursive, finite-dimensional solution to the (state) filtering problem, which, in the case of uncountable state spaces, is essentially limited to linear systems and the so-called Beneš class of systems \cite{benes1981}. However, as it turns out, the class of stochastic quantum systems studied in the present work also admits such a solution in the form of the quantum filter, making them prime candidates for application of methodologies relying on such filtering solutions. 




The rest of the paper is organized as follows. In Section \ref{sec:DT}, we present the quantum stochastic master equation in discrete-time and derive the online ML estimator. In Section \ref{sec:CT}, we use the considerations from the discrete-time formulation to derive the continuous-time online ML estimator. In Section \ref{sec:simulations}, we demonstrate the algorithm in simulations on a simple two-level system. Lastly, in Section \ref{sec:discussion}, we discuss our findings and possibilities for future work.

\section{Discrete-Time Formulation}\label{sec:DT}
In the following, \revised{we will consider quantum states in the form of density operators $\rho$ acting on a (finite-dimensional) Hilbert space $\mathcal{H}=\mathbb{C}^n$, i.e., the set $\mathbb{S}=\{\rho \in \mathbb{C}^{n\times n} \mid \rho \geq 0, \Tr(\rho)=1\}$}. Throughout the paper, we will use the tilde (\~{}) accent to denote estimated quantities.

In the discrete-time formalism, upon observing the classical measurement $y_k$, the quantum state $\rho_k$ evolves from time $k$ to $k+1$ according to
\begin{align}
	\rho_{k+1} &= \frac{\mathcal{K}_{y_{k},\theta} (\rho_k)}{\Tr(\mathcal{K}_{y_{k},\theta} (\rho_k))}	\label{eq:quantum_DT}
\end{align}
with probability
\begin{align}
	\mathbb{P}(y_k=y \mid \rho_k, \theta) &= \Tr(\mathcal{K}_{y,\theta} (\rho_k)),
\end{align}
where $\mathcal{K}_{y,\theta} (\rho)\triangleq \sum_\mu M^\theta_{y,\mu}\rho M_{y,\mu}^{\theta\dagger}$ is a partial Kraus map corresponding to measurement outcome $y$ that depends on parameter $\theta\in\mathbb{R}^p$. \revised{The form of the operators $M^\theta_{y,\mu}$ is arbitrary and may represent both measurements and other dynamics; the only condition is that the partial Kraus maps satisfy the completeness relation $\sum_y \Tr(\mathcal{K}_{y,\theta}(\rho)) = 1$. Additionally, we will assume each $M^\theta_{y,\mu}$ to be differentiable with respect to $\theta$.}

Based on the classical measurement signal $y_k$, we want to recursively compute the ML estimate $\tilde{\theta}$ of the parameter $\theta$. 
\subsection{Maximum Likelihood Estimation}
We define the (incomplete-data\footnote{\revised{The term \textit{incomplete data} is sometimes used to convey the fact that the true state $\rho_k$ is unknown. In quantum systems, this will always be the case.}}) likelihood function at time $T$ as
\begin{equation}
	\begin{aligned}
		\lkhood_T(\theta; y_{T}, \tilde{\rho}_0) &\triangleq \mathbb{P}(y_0,\dots,y_T\mid \theta, \tilde{\rho}_0) \\
		&= \Tr(\mathcal{K}_{y_{T},\theta} \circ \dots \circ \mathcal{K}_{y_{0},\theta}(\tilde{\rho}_0))\\
		&=\prod_{k=0}^T \Tr(\mathcal{K}_{y_{k},\theta}(\tilde{\rho}_k)),
	\end{aligned}\label{eq:likelihood_def}%
\end{equation}
where $\tilde{\rho}_k$ is the solution to the quantum filter
\begin{align}
	\tilde{\rho}_{k+1} = \frac{\mathcal{K}_{y_{k},\theta} (\tilde{\rho}_k)}{\Tr(\mathcal{K}_{y_{k},\theta} (\tilde{\rho}_k))}.\label{eq:quantum_filter_DT}
\end{align}
%
%
%
\revised{We emphasize that, since the true state $\rho_k$ is unknown, the likelihood function  $\tilde{\rho}$ is computed based on the best available estimate $\tilde{\rho}_k$ from the quantum filter.}

From \eqref{eq:likelihood_def}, it is clear that we can formulate the likelihood function recursively as
\begin{equation}
	\begin{aligned}
		\lkhood_{k+1} = \Tr(\mathcal{K}_{y_{k+1},\theta}(\tilde{\rho}_{k+1}))  \lkhood_k. 
	\end{aligned}
\end{equation}
Introducing the log-likelihood function as $\ell_k \triangleq \log\lkhood_k$, we get
\begin{align}
	\ell_{k+1} = \log \Tr(\mathcal{K}_{y_{k+1},\theta}(\tilde{\rho}_{k+1}))  + \ell_k.\label{eq:log_likelihood_recursion}
\end{align}

\subsubsection{Offline gradient ascent}
A common method for finding the parameter value that maximizes the log-likelihood is via gradient ascent, i.e., by updating the parameter estimate $\tilde{\theta}$ (from some initial guess $\tilde{\theta}_0)$ by stepping in the direction of the gradient with some (possibly varying) step size $\gamma_i>0$, i.e.,
\begin{align}
	\tilde{\theta}_{i+1} = \tilde{\theta}_i + \gamma_i\nabla_\theta \ell_T(\tilde{\theta}_i).
\end{align}
To be exact, the gradient $\nabla_\theta \ell_T(\tilde{\theta}_i)$ (evaluated at the $i$'th iteration of the parameter estimate) depends on the full history of observations $y_0,\dots,y_T$ at time $T$. Computing the (exact) gradient can be done in different ways, namely, via the smoothing distribution or via propagation of the so-called sensitivity equations. In this work, we only focus on the latter approach as this naturally lends itself to an eventual (approximate) online formulation. 


We first note that the gradient of the log-likelihood with respect to the $p$-dimensional parameter $\theta$ can be written as
\begin{align}
	\nabla_\theta \ell_{T} &=  \begin{bmatrix}
		\dpd{\ell_{T}}{\theta_1} \\ \vdots \\  \dpd{\ell_{T}}{\theta_p}
	\end{bmatrix} + \sum_{k=0}^T \begin{bmatrix}
		\dpd{\ell_{T}}{\tilde{\rho}_k} \cdot \dpd{\tilde{\rho}_k}{\theta_1} \\ \vdots \\  \dpd{\ell_{T}}{\tilde{\rho}_k} \cdot \dpd{\tilde{\rho}_k}{\theta_p}
	\end{bmatrix},\label{eq:gradient_DT_full}
\end{align}
where we define the filter sensitivity at time $k$ as $\xi_{j,k} \triangleq \pd{\tilde{\rho}_k}{\theta_j}$ \revised{and use the dot symbol to denote the (Hilbert-Schmidt) inner product between two operators $A,B$, i.e., $A\cdot B \triangleq \Tr(A^\dagger B)$}. \revised{We remark that $\xi_{j,k}$ belongs to the tangent space of density operators, thus being a traceless Hermitian operator of similar dimension as $\rho$}. For notational convenience, we will in the following only consider scalar $\theta$ and suppress the $j$-subscript, but at the end of the section, we will write the algorithm out in full for $p$-dimensional $\theta$.

Introducing the notation \revised{$\Delta\ell_k \triangleq \ell_{k}-\ell_{k-1} = \log \Tr(\mathcal{K}_{y_{k},\theta}(\tilde{\rho}_{k}))$}, we can write the gradient recursively as
\begin{equation}
\begin{aligned}
	\nabla_\theta \ell_{k} &= \sum_{j=0}^k \dpd{\Delta\ell_{j}}{\theta} + 
		\dpd{\Delta\ell_{j}}{\tilde{\rho}_j} \cdot \dpd{\tilde{\rho}_j}{\theta}\\
	&= \dpd{\Delta\ell_{k}}{\theta} + 
		\dpd{\Delta\ell_{k}}{\tilde{\rho}_k} \cdot \dpd{\tilde{\rho}_k}{\theta} + \nabla_\theta \ell_{k-1}\\
	&= \frac{\Tr(\partial_\theta\mathcal{K}_{y_{k},\theta}(\tilde{\rho}_k) + \mathcal{K}_{y_{k},\theta}(\xi_{k}))}{\Tr(\mathcal{K}_{y_{k},\theta}(\tilde{\rho}_{k}))} + \nabla_\theta \ell_{k-1},\label{eq:gradient_DT}
\end{aligned}
\end{equation}
%
%
where the filter sensitivity $\xi_{k} \triangleq \pd{\tilde{\rho}_k}{\theta}$ has its own dynamics, found by differentiation of the filter evolution \eqref{eq:quantum_filter_DT} with respect to $\theta$, i.e.,
\begin{equation}
\begin{aligned}
	\xi_{k+1} &= \dpd{\left(\frac{\mathcal{K}_{y_{k},\theta} (\tilde{\rho}_k)}{\Tr(\mathcal{K}_{y_{k},\theta} (\tilde{\rho}_k))}\right)}{\theta} + \dpd{\left(\frac{\mathcal{K}_{y_{k},\theta} (\tilde{\rho}_k)}{\Tr(\mathcal{K}_{y_{k},\theta} (\tilde{\rho}_k))}\right)}{\tilde{\rho}_k} \cdot \xi_{k} \\
	&=\frac{\partial_{\theta}\mathcal{K}_{y_{k},{\theta}}(\tilde{\rho}_k) +\mathcal{K}_{y_{k},{\theta}}(\xi_{k})}{\Tr(\mathcal{K}_{y_{k},{\theta}}(\tilde{\rho}_{k}))} \\ &\hspace{1em}-\frac{\Tr(\partial_{\theta}\mathcal{K}_{y_{k},{\theta}}(\tilde{\rho}_k) + \mathcal{K}_{y_{k},{\theta}}(\xi_{k}) ) \tfrac{\mathcal{K}_{y_{k},{\theta}}(\tilde{\rho}_k) }{\Tr(\mathcal{K}_{y_{k},{\theta}}(\tilde{\rho}_k))}}{\Tr(\mathcal{K}_{y_{k},{\theta}}(\tilde{\rho}_{k}))}.
\end{aligned}\label{eq:sensitivity_DT}%
\end{equation}

When these equations (i.e., the quantum filter \eqref{eq:quantum_filter_DT} and the filter sensitivity \eqref{eq:sensitivity_DT}) are simulated for a fixed parameter value $\theta$, the gradient \eqref{eq:gradient_DT} is exact and may readily be used in an offline gradient-based ML scheme. However, due to the large computational burden of simulating these equations from time $0$ to time $T$ for each parameter update step, we seek an online approximation to the parameter update that can easily incorporate new observations as they arrive and, in particular, allow tracking of slow changes in the true, underlying parameter. 

\subsubsection{Online gradient ascent}\label{sec:online_GA_DT}
To make the aforementioned offline algorithm work in an online fashion, we take the following approach: Instead of simulating the entire trajectory of the filter and sensitivity equations with a fixed parameter to compute the gradient, we approximate the gradient by only simulating the latest recursion before updating the parameter estimate. In other words, we perform an online gradient ascent update in the form of
\begin{align}
	\tilde{\theta}_{k+1} = \tilde{\theta}_k + \gamma_k\nabla_\theta \Delta \ell_k(\tilde{\theta}_k),\label{eq:online_GA_DT}
\end{align}
\revised{where we emphasize that the gradient is computed only for the latest difference of the log-likelihood $\Delta \ell_k$.} This approach may equivalently be interpreted as simply assuming that $\ell_k$ in the recursion for $\ell_{k+1}$ in \eqref{eq:log_likelihood_recursion} is independent of $\theta$ and $\rho_k$.

To accommodate the online parameter update, the next recursion of the quantum filter and the quantum filter sensitivity will be computed using the latest parameter estimate. Hence, the full online gradient ascent algorithm can be summarized as the following set of coupled difference equations (for $\theta\in\mathbb{R}^p$):%
\begin{subequations}
\begin{align}
	\tilde{\rho}_{k+1} &= \frac{\mathcal{K}_{y_{k},\tilde{\theta}_k} (\tilde{\rho}_k)}{\Tr(\mathcal{K}_{y_{k},\tilde{\theta}_k}(\tilde{\rho}_{k}))},\\
	\xi_{j,k+1} &=\frac{\partial_{\theta_j}\mathcal{K}_{y_{k},\tilde{\theta}_k}(\tilde{\rho}_k) +\mathcal{K}_{y_{k},\tilde{\theta}_k}(\xi_{j,k})}{\Tr(\mathcal{K}_{y_{k},\tilde{\theta}_k}(\tilde{\rho}_{k}))}\nonumber\\ -&\frac{\Tr(\partial_{\theta_j}\mathcal{K}_{y_{k},\tilde{\theta}_{k}}(\tilde{\rho}_k) + \mathcal{K}_{y_{k},\tilde{\theta}_{k}}(\xi_{j,k}) ) \tfrac{\mathcal{K}_{y_{k},\tilde{\theta}_k}(\tilde{\rho}_k) }{\Tr(\mathcal{K}_{y_{k},\tilde{\theta}_k}(\tilde{\rho}_k))}}{\Tr(\mathcal{K}_{y_{k},\tilde{\theta}_k}(\tilde{\rho}_{k}))},\label{eq:full_online_GA_DTb}\\
	\tilde{\theta}_{j,k+1} &=
	\tilde{\theta}_{j,k} + \gamma_k\ \frac{\Tr(\partial_{\theta_j}\mathcal{K}_{y_{k},\tilde{\theta}_k}(\tilde{\rho}_k) + \mathcal{K}_{y_{k},\tilde{\theta}_k}(\xi_{j,k}))}{\Tr(\mathcal{K}_{y_{k},\tilde{\theta}_k}(\tilde{\rho}_{k}))}\label{eq:full_online_GA_DTc}%
\end{align}\label{eq:full_online_GA_DT}%
\end{subequations}%
for $j=1,\dots,p$. 

\section{Continuous-Time Formulation}\label{sec:CT}
In this section, we will consider the diffusive quantum stochastic master equation given by the following stochastic differential equation (SDE), to be understood in the sense of Itô: 
\begin{subequations}
	\begin{align}
		\dif \rho_t &= -i\comm{H(\theta)}{\rho}\dif t + \mathcal{D}[L](\rho_t) \dif t + \sqrt{\eta}\mathcal{H}[L](\rho_t)\dif W_t,\label{eq:quantum_SME} \\
		\dif y_t &= \sqrt{\eta}\Tr((L+L^\dagger)\rho_t) \dif t + \dif W_t\label{eq:dy}
	\end{align}\label{eq:quantum_sys}%
\end{subequations}%
with the superoperators $\mathcal{D}[L](\rho)$ and $\mathcal{H}[L](\rho)$ defined as
\begin{align*}
	\mathcal{D}[L](\rho) &\triangleq L\rho L^\dagger -\frac{1}{2}(L^\dagger L \rho +  \rho L^\dagger L),\\
	\mathcal{H}[L](\rho) &\triangleq L\rho + \rho L^\dagger - \Tr((L+L^\dagger)\rho)\rho.
\end{align*}
Here, $H(\theta)$ (a Hermitian operator that depends on parameter $\theta$) denotes the Hamiltonian of the system. \revised{The system is subject to a continuous measurement defined in terms of the (not necessarily Hermitian) operator $L$ (typically called a jump operator or Lindblad operator) and the detection efficiency $\eta\in[0,1]$ with $y_t$ being the classical measurement signal and $W_t$ being a standard Wiener process. The term $\mathcal{D}[L](\rho)$ corresponds to the measurement-induced decoherence, whereas the term $\mathcal{H}[L](\rho)$ models the (conditional) measurement back-action.} Note that the same Wiener process is driving both the process and the measurement signal---a unique feature of quantum systems---and that for $\eta=0$, the measurement signal contains no information about the quantum state. For more details on the properties of the stochastic master equation, we refer to \cite{wiseman2010,rouchon2022} and the references therein. 

\revised{For simplicity of presentation, we only consider the case where there is just a single measurement channel (i.e., a single jump operator $L$) with corresponding decoherence, but the model may easily be extended to the case where there are several measurement and decoherence channels.} Likewise, we only consider the case where the parameter $\theta$ enters through the Hamiltonian, although we relax this assumption in the simulation examples in Section \ref{sec:simulations}.

In the following, we derive an online gradient ascent algorithm in continuous-time for computing the ML estimate $\tilde{\theta}$ of the parameter $\theta$.


With standard Itô rules (i.e., $\dif W_t^2=\dif t$, $\dif t^2=0$ and $\dif W_t\! \dif t=0$), the system \eqref{eq:quantum_sys} admits the equivalent formulation\footnote{To be interpreted in the sense $\dif \rho_t = \rho_{t+\dif t}-\rho_t$.}
\begin{align}
	\rho_{t+\dif t} = \frac{M_{\dif y_{t}}^\theta \rho M_{\dif y_{t}}^{\theta\dagger} + (1-\eta) L\rho_t L^\dagger \dif t}{\Tr(M_{\dif y_{t}}^\theta \rho M_{\dif y_{t}}^{\theta\dagger} + (1-\eta) L\rho_t L^\dagger \dif t)}\label{eq:sde_kraus}
\end{align}
with $M_{\dif y_{t}}^\theta \triangleq I -(iH(\theta) + \frac{1}{2}L^\dagger L)\dif t + \sqrt{\eta} L \dif y_{t}$, where we recognize that this is a Kraus map formulation similar to \eqref{eq:quantum_DT}. Writing out the partial Karus map in full, we have, still under the Itô interpretation,
\begin{equation}
	\begin{aligned}
		\mathcal{K}_{\dif y,\theta}(\rho) &= M_{\dif y}^\theta \rho M_{\dif y}^{\theta\dagger} + (1-\eta) L\rho L^\dagger \dif t \\
&= \rho - i\comm{H(\theta)}{\rho}\dif t + \mathcal{D}[L](\rho)\dif t\\
& \hspace*{8em}+ \sqrt{\eta}(L\rho + \rho L^\dagger)\dif y.
	\end{aligned}\label{eq:K_dy}
\end{equation}
%
%
\begin{figure*}[!b]
	\normalsize
	\setcounter{mytempeqncnt}{\value{equation}}
	\setcounter{equation}{22}
	\hrulefill
	\vspace*{-2.3pt}
	\begin{subequations}
		\begin{align}
			\dif \tilde{\rho}_t &= \left(-i[H(\tilde{\theta}_t),\tilde{\rho}_t] + \mathcal{D}[L](\tilde{\rho}_t)\right) \dif t + \sqrt{\eta}
			\mathcal{H}[L](\tilde{\rho}_t) \left(\dif y_t - \sqrt{\eta}\Tr((L+L^\dagger)\tilde{\rho}_t)\dif t\right) \\
			\dif \xi_{j,t} &= \left(-i\left[\partial_{\theta_j} H(\tilde{\theta}_t),\tilde{\rho}_t\right] -i[H(\tilde{\theta}_t),\xi_{j,t}] + \mathcal{D}[L](\xi_{j,t})  -\eta\mathcal{H}[L](\tilde{\rho}_t)\Tr((L+L^\dagger)\xi_{j,t})\right) \dif t \notag \\
			&\hspace*{1em}+ \sqrt{\eta} \left(L\xi_{j,t} + \xi_{j,t} L^\dagger - \left(\Tr((L+L^\dagger)\xi_{j,t})\tilde{\rho}_t + \Tr((L+L^\dagger)\tilde{\rho}_t)\xi_{j,t}\right)\right) \left(\dif y_t - \sqrt{\eta}\Tr((L+L^\dagger)\tilde{\rho}_t)\dif t\right) \\
			\dif \tilde{\theta}_{j,t} &=  \gamma_t \sqrt{\eta} \Tr((L+L^\dagger)\xi_{j,t}) \left(\dif y_t - \sqrt{\eta}\Tr((L+L^\dagger)\tilde{\rho}_t)\dif t\right)
		\end{align}\label{eq:full_filter_CT}%
	\end{subequations}
%
	\setcounter{equation}{\value{mytempeqncnt}}
\end{figure*}
From \eqref{eq:likelihood_def}, the likelihood function may now be written as
\begin{equation}
	\begin{aligned}
		\lkhood_{t+\dif t} &= 
		\lkhood_{t+\dif t}(\theta; y_{t+\dif t}, \tilde{\rho}_0) 
		\\
		&=\Tr(\mathcal{K}_{\dif y_{t},\theta}(\tilde{\rho}_t)) \lkhood_{t}  \\
		&= \left(1 + \sqrt{\eta}\Tr((L+L^\dagger)\tilde{\rho}_t)\dif y_t\right) \lkhood_{t} 
	\end{aligned}
\end{equation}
from which the log-likelihood immediately follows to be\footnote{Recall the power series expansion $\log x = \sum_{k=1}^\infty (-1)^{k+1} \frac{(x-1)^k}{k}$.}
\begin{multline}
	\ell_{t+\dif t} =\sqrt{\eta}\Tr((L+L^\dagger)\tilde{\rho}_t) \\
	\times\left(\dif y_t -\frac{\sqrt{\eta}}{2}\Tr((L+L^\dagger)\tilde{\rho}_t)\dif t\right) +\ell_{t}.
\end{multline}
The full log-likelihood may then be written in integral form as
\begin{multline}
	\ell_t(\theta; y_t,\tilde{\rho}_0) = \sqrt{\eta}\int_0^t \Tr((L+L^\dagger)\tilde{\rho}_s) \dif y_s \\ 
	- \frac{\eta}{2} \int_0^t \Tr((L+L^\dagger)\tilde{\rho}_s)^2 \dif s,\label{eq:log_likelihood_CT}
\end{multline}
where $\tilde{\rho}_t$ is the solution to the quantum filter
\begin{multline}
	\dif \tilde{\rho}_t = -i[H({\theta}),\tilde{\rho}_t]\dif t + \mathcal{D}[L](\tilde{\rho}_t) \dif t \\
	+ \sqrt{\eta}\mathcal{H}[L](\tilde{\rho}_t)\left( \dif y_t - \sqrt{\eta}\Tr((L+L^\dagger)\tilde{\rho}_t)\dif t\right).\label{eq:quantum_filter}
\end{multline}

Following the same reasoning as in Section \ref{sec:online_GA_DT}, we will design a continuous-time online gradient ascent update for finding the ML parameter estimate $\tilde{\theta}$ (from some initial guess $\tilde{\theta}_0)$. Hence, analogously to the discrete-time online parameter update \eqref{eq:online_GA_DT}, we have
\begin{equation}
\begin{aligned}
	\dif \tilde{\theta}_t &= \gamma_t \nabla_\theta \dif \ell_t(\tilde{\theta}_t)\\
	&=\gamma_t \nabla_\theta \Bigg(\sqrt{\eta}\Tr((L+L^\dagger)\tilde{\rho}_t) \\
	&\hspace*{4em}\times\left(\dif y_t -\frac{\sqrt{\eta}}{2}\Tr((L+L^\dagger)\tilde{\rho}_t)\dif t\right)\Bigg)
\end{aligned}\label{eq:dtheta}
\end{equation}
with $\dif \ell_t$ denoting the integrand of \eqref{eq:log_likelihood_CT} and $\gamma_t>0$ being a (possibly time-varying) learning rate. We note here that \eqref{eq:dtheta} may be meaningfully treated as an Itô process as $\ell_t$ is itself an Itô process \cite{lipster2001a}. 

Formally, we write the gradient of the log-likelihood with respect to the parameter $\theta$ as
\begin{align}
	\nabla_\theta \dif \ell_t = \dpd{\dif \ell_t}{\theta} + \dpd{\dif \ell_t}{\tilde{\rho}_t} \cdot \dpd{\tilde{\rho}_t}{\theta},
\end{align}
where the filter sensitivity $\xi_t \triangleq \pd{\tilde{\rho}_t}{\theta}$ \revised{(defined analogously to the discrete-time case \eqref{eq:gradient_DT_full})} satisfies the sensitivity equation
\begin{align}
	\dif \xi_t = \dpd{\dif \tilde{\rho}_t}{\theta} + \dpd{\dif \tilde{\rho}_t}{\tilde{\rho}_t} \cdot \xi_t
\end{align}
with $\dif \tilde{\rho}_t$ denoting the quantum filtering equation \eqref{eq:quantum_filter}. \revised{Writing out the sensitivity equation and the parameter update out in full for $p$-dimensional $\theta$, we get the diffusion process \eqref{eq:full_filter_CT} for $j=1,\dots,p$ (placed at the bottom of this page).}

We note here that all three equations (i.e., the quantum filter, the filter sensitivity and the parameter update) are driven by the same innovation process $\dif I_t \triangleq \left(\dif y_t - \sqrt{\eta}\Tr((L+L^\dagger)\tilde{\rho}_t)\dif t\right)$. From this formulation, it is also clear that the learning rate $\gamma_t$ will directly amplify the effect of the noise inherent to the innovation process on the parameter estimate. Hence, in order for the noise to be suppressed, the learning rate must be chosen to be sufficiently slow. From a slightly different point of view, the need for a slow learning rate may also be argued as follows: 

\revised{Assuming the system to be ergodic\footnote{\revised{Ergodic in the sense that the system has a unique invariant measure. This is typically a necessary condition for stochastic online algorithms to converge (see, e.g., \cite{benveniste1990,borkar2022}).}}, the innovation process must, for a fixed parameter value, reach a stationary distribution}, which must consequently imply that both the state estimate $\tilde{\rho}_t$ and the filter sensitivity $\xi_t$ have reached a stationary distribution, and from this stationary distribution, the gradient must be perceptible. Hence, if the change in the parameter estimate $\tilde{\theta}_t$ is too fast, the stochastic estimate of the gradient will not have time to stabilize and the perceived gradient will instead be dominated by random noise.

Before proceeding, we remark that the set of equations \eqref{eq:full_filter_CT} could equivalently be derived directly from \eqref{eq:full_online_GA_DT} via the Kraus map \eqref{eq:sde_kraus} and the Itô rule. We may thus interpret \eqref{eq:full_online_GA_DT} as a (first-order) numerical integration scheme of the continuous-time online gradient ascent \eqref{eq:full_filter_CT}. In fact, when \eqref{eq:full_filter_CT} has to be simulated, we will always recommend using the discrete-time equivalent \eqref{eq:full_online_GA_DT} as opposed to, e.g., a naive application of the Euler-Maruyama scheme, in order to properly preserve positivity and trace\footnote{The discretization scheme here is not exactly trace-preserving, but this can be handled by a small adjustment presented in \cite[Section~3.4]{rouchon2022}.} of the quantum state.

\section{Simulation Example}\label{sec:simulations}

In the following, we will consider a two-level quantum system undergoing homodyne measurement, described by the stochastic master equation \eqref{eq:quantum_sys} with Hamiltonian and Lindblad operator given by
\begin{align}\addtocounter{equation}{1}
	H = \frac{\Delta}{2}\sigma_z + \frac{\Omega}{2}\sigma_x, \qquad L=\sqrt{\kappa} \sigma_z,\label{eq:sim_example}
\end{align}
where $\sigma_{j\in\{x,y,z\}}$ denotes the Pauli matrices.
	
This system could, for instance, represent a superconducting qubit in a rotating wave approximation, in which case $\Delta$ is the detuning between the qubit frequency and the driving field, and $\Omega$ is the Rabi frequency \cite{blais2021}. We typically have that these two parameters are of the same order of magnitude and much greater than the measurement rate $\kappa$, i.e., $\kappa \ll \abs{\Delta}$, $\abs{\Omega}$.

As discussed as the end of Section \ref{sec:CT}, the learning rate $\gamma_t$ must be sufficiently slow compared to the dynamics of the rest of the system. For the example at hand, the measurement rate $\kappa$ dictates both the rate of information extraction as well as the slowest dynamics in the system. Hence, we pick $\gamma = \frac{\kappa}{1000}$. 

In the following, we will estimate all four parameters (i.e., the frequencies $\Omega$ and $\Delta$, the detection efficiency $\eta$ and the measurement rate $\kappa$) simultaneously along a single quantum trajectory. 

\begin{figure*}[ht]
	\begin{tabular}{lcr}
		\begin{tabular}[t]{l}
	\subfloat[Estimation of static parameters. The true parameter values (dashed lines) are $(\Omega,\eta,\Delta,\kappa)=(1, 0.7, 0.2, 0.1)$. The estimates (solid lines) are initialized at $(\tilde{\Omega}_0, \tilde{\eta}_0, \tilde{\Delta}_0, \tilde{\kappa}_0)=(1.3, 0.6, 0.3, 0.15)$.\label{fig:static}]{	\tikzsetnextfilename{claus1a}
	\begin{tikzpicture}[trim axis left, trim axis right]
		\def\gam{0.0001}
		\begin{groupplot}[group style={group size=2 by 1, horizontal sep=1cm, vertical sep=4cm, xlabels at=edge bottom},default, wide3,enlargelimits=false, legend pos=outer north east, xmin=0, ymin=0,]
			
			\nextgroupplot[
			ymax=1.6,ymin=0, xlabel=Time $\gamma t$, legend image post style={scale=0.6},
			legend columns=4,
			transpose legend=true,
			legend entries={${\Omega}$, $\eta$, $\Delta$,$\kappa$,\vphantom{$\Omega$}\hspace*{-0.1em},\vphantom{$\eta$}\hspace*{-0.1em},\vphantom{$\Delta$}\hspace*{-0.1em},\vphantom{$\kappa$}\hspace*{-0.1em}},
			legend to name=legend1,
			]
			\addplot+[] table[x expr=\thisrowno{0}*\gam] {../../../Matlab/plotdata/online_maximum_likelihood_4par_sim1/mu.dat};
			
			\addplot+[] table[x expr=\thisrowno{0}*\gam] {../../../Matlab/plotdata/online_maximum_likelihood_4par_sim1/eta.dat};

			\addplot+[] table[x expr=\thisrowno{0}*\gam] {../../../Matlab/plotdata/online_maximum_likelihood_4par_sim1/Delta.dat};
			
			\addplot+[] table[x expr=\thisrowno{0}*\gam] {../../../Matlab/plotdata/online_maximum_likelihood_4par_sim1/kappa.dat};
			
			\pgfplotsset{cycle list shift=-4}
			
			\addplot+[dashed] table [row sep=\\] {
				X Y\\
				0 1\\
				20 1\\
			};
		
			\addplot+[dashed] table [row sep=\\] {
				X Y\\
				0 0.7\\
				20 0.7\\
			};
		
			\addplot+[dashed] table [row sep=\\] {
				X Y\\
				0 0.2\\
				20 0.2\\
			};
		
			\addplot+[dashed] table [row sep=\\] {
				X Y\\
				0 0.1\\
				20 0.1\\
			};

		\end{groupplot}
	\end{tikzpicture}
}	\end{tabular}
	& \hspace*{-0.8em}\begin{tabular}[b]{c}
	\tikzsetnextfilename{claus1legend}\begin{tikzpicture}
		\tikzset{%
			fuckaitor/.style    = {anchor=west,black,minimum width=1.5em,outer sep=0pt},
		}
		
		\draw[Dark2-A,thick] (0,0) to ++(1em,0) node[fuckaitor](omega){$\Omega$};
		\draw[Dark2-A,thick,dashed] (omega.east) to ++(1em,0);
		
		\draw[Dark2-B,thick] (0,-1em) to ++(1em,0) node[fuckaitor](eta){$\eta$};
		\draw[Dark2-B,thick,dashed] (eta.east) to ++(1em,0);
		
		\draw[Dark2-C,thick] (0,-2em) to ++(1em,0) node[fuckaitor](delta){$\Delta$};
		\draw[Dark2-C,thick,dashed] (delta.east) to ++(1em,0);
		
		\draw[Dark2-D,thick] (0,-3em) to ++(1em,0) node[fuckaitor](kappa){$\kappa$};
		\draw[Dark2-D,thick,dashed] (kappa.east) to ++(1em,0) node[](se){};
		
		\draw[draw=black] (-0.5em,0.8em) rectangle ($(se)+(0.5em,-0.8em)$);	
	\end{tikzpicture}\vspace{1.5cm}
	\end{tabular}
	&\hspace*{-1.16em}\begin{tabular}[t]{r}
	\subfloat[\revised{Tracking of slow-varying parameters. The true parameter values (dashed lines) are given by the static parameter values from Fig. \ref{fig:static} plus a time-dependent perturbation $(\Omega',\eta',\Delta',\kappa')_t=\big(0.5\sin(0.12\gamma t),$ $0.2\sin(0.05\gamma t),$ $0.16\sin(0.12 \gamma t),$ $0.1\sin(0.11\gamma t)\big)$. The estimates (solid lines) are initialized at $(\tilde{\Omega}_0, \tilde{\eta}_0, \tilde{\Delta}_0, \tilde{\kappa}_0)=(1.3, 0.8, 0.3, 0.15)$.} %
	\label{fig:varying}]{		\tikzsetnextfilename{claus1b}
	\begin{tikzpicture}[trim axis left, trim axis right]
		\def\gam{0.0001}
		\begin{groupplot}[group style={group size=2 by 1, horizontal sep=1cm, vertical sep=4cm, xlabels at=edge bottom},default,wide3,enlargelimits=false,legend pos=north east, xmin=0, ymin=0,yticklabel pos=right,]
			
			\nextgroupplot[
			ymax=1.6,ymin=0, legend columns=3, xlabel=Time $\gamma t$]
			\addplot+[] table[x expr=\thisrowno{0}*\gam] {../../../Matlab/plotdata/online_maximum_likelihood_4par_sim1_varying_wrong_init1/mu.dat};
			\addplot+[] table[x expr=\thisrowno{0}*\gam] {../../../Matlab/plotdata/online_maximum_likelihood_4par_sim1_varying_wrong_init1/eta.dat};
			\addplot+[] table[x expr=\thisrowno{0}*\gam] {../../../Matlab/plotdata/online_maximum_likelihood_4par_sim1_varying_wrong_init1/Delta.dat};
			\addplot+[] table[x expr=\thisrowno{0}*\gam] {../../../Matlab/plotdata/online_maximum_likelihood_4par_sim1_varying_wrong_init1/kappa.dat};
			\pgfplotsset{cycle list shift=-4}

			\addplot+[dashed] table[x expr=\thisrowno{0}*\gam] {../../../Matlab/plotdata/online_maximum_likelihood_4par_sim1_varying_wrong_init1/mu_t.dat};
			\addplot+[dashed] table[x expr=\thisrowno{0}*\gam] 	{../../../Matlab/plotdata/online_maximum_likelihood_4par_sim1_varying_wrong_init1/eta_t.dat};
			\addplot+[dashed] table[x expr=\thisrowno{0}*\gam] 	{../../../Matlab/plotdata/online_maximum_likelihood_4par_sim1_varying_wrong_init1/Delta_t.dat};
			\addplot+[dashed] table[x expr=\thisrowno{0}*\gam] 	{../../../Matlab/plotdata/online_maximum_likelihood_4par_sim1_varying_wrong_init1/kappa_t.dat};


			
			
		\end{groupplot}

	\end{tikzpicture}%
}
		\end{tabular}%
\end{tabular}
	\caption{Two examples of sample trajectories for the parameter estimates for the two-level quantum system described by the diffusive master equation \eqref{eq:quantum_sys} with Hamiltonian and Lindblad operator given by \eqref{eq:sim_example}. The learning rate in both examples is constant and given by $\gamma=\frac{\kappa_0}{1000}=10^{-4}$. Simulations are done using the discrete-time model \eqref{eq:full_online_GA_DT} and Kraus map \eqref{eq:sde_kraus} with a time-discretization of $\Delta t=10^{-2}$. The solid lines correspond to the estimated parameters, whereas the dashed lines correspond to the true parameter values.}\label{fig:simulation}	
\end{figure*}
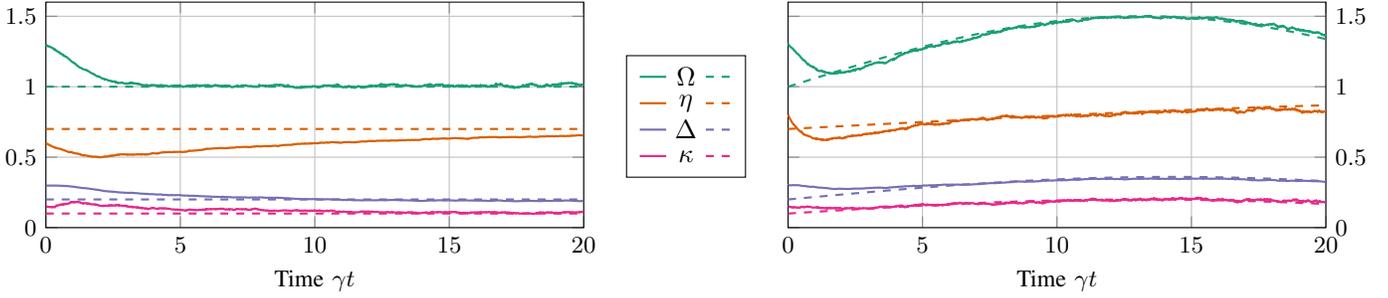

\revised{We consider two situations: (i) The parameters are static (see Fig.\ \ref{fig:static}), and (ii) the parameters vary slowly over time (see Fig.\ \ref{fig:varying}). In both examples, the parameter estimates are initialized with values different from the true parameter values.}

In the simulations, the true initial state is the ground state $\rho_0=\op{0}$, the initial filter state is the maximally mixed state $\tilde{\rho}_0=\frac{1}{2}I$, and the filter sensitivities are initialized at $\xi_{j,0}=0$ for $j\in\{\Omega,\Delta,\eta,\kappa\}$. The true parameter values and the initial parameter estimates may be found from Fig. \ref{fig:simulation}. The simulations are performed using the discrete-time model \eqref{eq:full_online_GA_DT} and Kraus map \eqref{eq:sde_kraus} with a time-discretization of $\Delta t=10^{-2}$. Both simulations are run for $T=\frac{20}{\gamma}=200\cdot 10^{3}$ time units. For parameters in the MHz range, as would be typical for superconducting devices, this corresponds to 200 ms of data.

\revised{In terms of implementation of the algorithm, we estimate the quantities $\sqrt{\eta}$ and $\sqrt{\kappa}$ directly (as opposed to just $\eta$ and $\kappa$) for better numerical stability when computing the gradient.}

As seen from Fig.\ \ref{fig:static}, when the parameters are constant, the parameter estimates all eventually converge to the true values, although we note that $\eta$ converges significantly slower than the remaining parameters. In particular, we note that, for this specific example, the estimator for $\eta$ seems to have a behavior reminiscent of a non-minimum phase system in the sense that it initially undershoots with reference to the true value. 

Likewise, when the parameters are changing slowly over time, as seen in Fig.\ \ref{fig:varying}, the estimator is, to a large degree, able to track this change. A natural limitation in such tracking situations is the rate of information extraction (in this case quantified in terms of $\kappa$) and, consequently, the learning rate $\gamma$. A higher learning rate will in principle allow the estimator to react to faster changes in the parameters, but at the cost of estimation accuracy, as the larger learning rate will also amplify the unavoidable noise related to the quantum measurement process. 





%


%
%
%
%
%

\section{Discussion and Future Work}\label{sec:discussion}

In this work, we have proposed an online algorithm for computing the maximum likelihood (ML) estimate of unknown parameters in a continuously monitored quantum system and demonstrated its usefulness via a simple simulation example. 

For the particular example, the algorithm is shown to be able to estimate and track multiple slowly varying parameters based only on the observed measurement history and without relying on any dynamical model describing the time-evolution of the parameters.

A fundamental aspect to highlight for the reason for the proposed online ML algorithm to work is the notable feature of stochastic quantum systems that they admit a tractable (finite-dimensional\footnote{This is only the case when the Hilbert space on which the quantum state is defined is finite-dimensional.}), exact solution to the (state) filtering problem in the form of \eqref{eq:quantum_filter_DT} and \eqref{eq:quantum_filter} for, respectively, the discrete-time and continuous-time case. Hence, in the language of nonlinear filtering theory, we may consider quantum systems as part of the class of stochastic systems that admit a tractable solution to the Kushner-Stratonovich equation.


An important next step is to formally proof convergence (in a suitable sense) of the proposed algorithm. Typically, a necessary condition for online algorithms to converge is that the underlying dynamical system, for a fixed parameter $\theta$, converges to a unique invariant measure. In our case, this would entail finding conditions under which the true quantum system $\rho_t$, the corresponding quantum filter $\tilde{\rho}_t$ and its sensitivity $\xi_t$ all converge to a unique invariant measure. For the quantum filter, this can be ensured under a purification condition on the measurement operators and an ergodicity condition on the Lindbladian \cite{benoist2021,benoist2023}, whereas the sensitivity operator has yet to see any attention in the literature. We note, however, that simulations suggest that the purification condition (which automatically entails unit detection efficiency) is overly restrictive. When such ergodicity conditions have been established, it appears reasonable to approach the convergence proof in a way similar to that taken by \cite{surace2019}.

While such proofs, broadly based on stochastic approximation techniques \cite{borkar2022}, usually rely on an assumption on the learning rate $\gamma_t$ tending to 0 as time $t$ tends to infinity, practical applications, especially for tracking, usually requires a non-zero learning rate for all time. Hence, it is of great interest to derive bounds on the uncertainty of the parameter estimate given in terms of the magnitude of the learning rate.

As gradient descent is known to have a slow convergence rate, it might be fruitful to investigate more advanced parameter update rules such as (quasi-)Newton methods involving (approximate) second-order information or adaptive learning rates known from the (online) optimization literature \cite{hazan2022}. This is further motivated by the presented example, where it is clear that the induced dynamics can result in undesirable properties such as an initial undershoot in some parameter estimates.

Lastly, future work may include applying the proposed method to systems subject to jump-type measurements like photodetectors as well as infinite-dimensional systems such as systems involving bosonic modes.


\bibliography{bibtex/literature.bib}

\begin{thebibliography}{10}
\providecommand{\url}[1]{#1}
\csname url@samestyle\endcsname
\providecommand{\newblock}{\relax}
\providecommand{\bibinfo}[2]{#2}
\providecommand{\BIBentrySTDinterwordspacing}{\spaceskip=0pt\relax}
\providecommand{\BIBentryALTinterwordstretchfactor}{4}
\providecommand{\BIBentryALTinterwordspacing}{\spaceskip=\fontdimen2\font plus
\BIBentryALTinterwordstretchfactor\fontdimen3\font minus
  \fontdimen4\font\relax}
\providecommand{\BIBforeignlanguage}[2]{{%
\expandafter\ifx\csname l@#1\endcsname\relax
\typeout{** WARNING: IEEEtran.bst: No hyphenation pattern has been}%
\typeout{** loaded for the language `#1'. Using the pattern for}%
\typeout{** the default language instead.}%
\else
\language=\csname l@#1\endcsname
\fi
#2}}
\providecommand{\BIBdecl}{\relax}
\BIBdecl

\bibitem{wiseman2010}
H.~M. Wiseman and G.~J. Milburn, \emph{Quantum Measurement and Control}.\hskip
  1em plus 0.5em minus 0.4em\relax Cambridge University Press, 2010.

\bibitem{rouchon2022}
P.~Rouchon, ``A tutorial introduction to quantum stochastic master equations
  based on the qubit/photon system,'' \emph{Annual Reviews in Control},
  vol.~54, pp. 252--261, 2022.

\bibitem{blais2021}
A.~Blais, A.~L. Grimsmo, S.~Girvin, and A.~Wallraff, ``Circuit quantum
  electrodynamics,'' \emph{Reviews of Modern Physics}, vol.~93, no.~2, p.
  025005, May 2021.

\bibitem{amoros-binefa2021}
J.~Amorós-Binefa and J.~Kołodyński, ``Noisy atomic magnetometry in real
  time,'' \emph{New Journal of Physics}, vol.~23, no.~12, p. 123030, Dec. 2021.

\bibitem{rossi2019}
M.~Rossi, D.~Mason, J.~Chen, and A.~Schliesser, ``Observing and verifying the
  quantum trajectory of a mechanical resonator,'' \emph{Physical Review
  Letters}, vol. 123, no.~16, p. 163601, Oct. 2019.

\bibitem{jazwinski}
A.~H. Jazwinski, \emph{Stochastic Processes and Filtering Theory}.\hskip 1em
  plus 0.5em minus 0.4em\relax Academic Press, 1970.

\bibitem{verstraete2001}
F.~Verstraete, A.~C. Doherty, and H.~Mabuchi, ``Sensitivity optimization in
  quantum parameter estimation,'' \emph{Physical Review A}, vol.~64, no.~3, p.
  032111, Aug. 2001.

\bibitem{geremia2003}
J.~Geremia, J.~K. Stockton, A.~C. Doherty, and H.~Mabuchi, ``Quantum {Kalman}
  filtering and the {Heisenberg} limit in atomic magnetometry,'' \emph{Physical
  Review Letters}, vol.~91, no.~25, p. 250801, Dec. 2003.

\bibitem{stockton2004}
J.~K. Stockton, J.~M. Geremia, A.~C. Doherty, and H.~Mabuchi, ``Robust quantum
  parameter estimation: {Coherent} magnetometry with feedback,'' \emph{Physical
  Review A}, vol.~69, no.~3, p. 032109, Mar. 2004.

\bibitem{gambetta2001}
J.~Gambetta and H.~M. Wiseman, ``State and dynamical parameter estimation for
  open quantum systems,'' \emph{Physical Review A}, vol.~64, no.~4, p. 042105,
  Sep. 2001.

\bibitem{warszawski2004}
P.~Warszawski, J.~Gambetta, and H.~M. Wiseman, ``Dynamical parameter estimation
  using realistic photodetection,'' \emph{Physical Review A}, vol.~69, no.~4,
  p. 042104, Apr. 2004.

\bibitem{gammelmark2013}
S.~Gammelmark and K.~Mølmer, ``Bayesian parameter inference from continuously
  monitored quantum systems,'' \emph{Physical Review A}, vol.~87, no.~3, p.
  032115, Mar. 2013.

\bibitem{kiilerich2016}
A.~H. Kiilerich and K.~Mølmer, ``Bayesian parameter estimation by continuous
  homodyne detection,'' \emph{Physical Review A}, vol.~94, no.~3, p. 032103,
  Sep. 2016.

\bibitem{chase2009}
B.~A. Chase and J.~M. Geremia, ``Single-shot parameter estimation via
  continuous quantum measurement,'' \emph{Physical Review A}, vol.~79, no.~2,
  p. 022314, Feb. 2009.

\bibitem{negretti2013}
A.~Negretti and K.~Mølmer, ``Estimation of classical parameters via continuous
  probing of complementary quantum observables,'' \emph{New Journal of
  Physics}, vol.~15, no.~12, p. 125002, Dec. 2013.

\bibitem{six2015}
P.~Six, P.~Campagne-Ibarcq, L.~Bretheau, B.~Huard, and P.~Rouchon, ``Parameter
  estimation from measurements along quantum trajectories,'' in \emph{2015 54th
  {IEEE} {Conference} on {Decision} and {Control} ({CDC})}, Dec. 2015, pp.
  7742--7748.

\bibitem{ralph2017}
J.~F. Ralph, S.~Maskell, and K.~Jacobs, ``Multiparameter estimation along
  quantum trajectories with sequential {Monte} {Carlo} methods,''
  \emph{Physical Review A}, vol.~96, no.~5, p. 052306, Nov. 2017.

\bibitem{bompais2022}
M.~Bompais, N.~H. Amini, and C.~Pellegrini, ``Parameter estimation for quantum
  trajectories: {Convergence} result,'' in \emph{2022 {IEEE} 61st {Conference}
  on {Decision} and {Control} ({CDC})}, Dec. 2022, pp. 5161--5166.

\bibitem{ralph2011}
J.~F. Ralph, K.~Jacobs, and C.~D. Hill, ``Frequency tracking and parameter
  estimation for robust quantum state estimation,'' \emph{Physical Review A},
  vol.~84, no.~5, p. 052119, Nov. 2011.

\bibitem{cortez2017}
L.~Cortez, A.~Chantasri, L.~P. García-Pintos, J.~Dressel, and A.~N. Jordan,
  ``Rapid estimation of drifting parameters in continuously measured quantum
  systems,'' \emph{Physical Review A}, vol.~95, no.~1, p. 012314, Jan. 2017.

\bibitem{six2016}
P.~Six, P.~Campagne-Ibarcq, I.~Dotsenko, A.~Sarlette, B.~Huard, and P.~Rouchon,
  ``Quantum state tomography with noninstantaneous measurements, imperfections,
  and decoherence,'' \emph{Physical Review A}, vol.~93, no.~1, p. 012109, Jan.
  2016.

\bibitem{cappe2005}
O.~Cappé, E.~Moulines, and T.~Rydén, \emph{Inference in hidden {Markov}
  models}, ser. Springer series in statistics.\hskip 1em plus 0.5em minus
  0.4em\relax New York: Springer, 2005.

\bibitem{benveniste1990}
A.~Benveniste, M.~Métivier, and P.~Priouret, \emph{Adaptive {Algorithms} and
  {Stochastic} {Approximations}}.\hskip 1em plus 0.5em minus 0.4em\relax
  Springer Berlin Heidelberg, 1990.

\bibitem{ljung1983}
L.~Ljung and T.~Söderström, \emph{\BIBforeignlanguage{eng}{Theory and
  Practice of Recursive Identification}}, 1st~ed., ser. The {MIT} {Press}
  series in signal processing, optimization, and control.\hskip 1em plus 0.5em
  minus 0.4em\relax The MIT Press, 1983, no.~4.

\bibitem{surace2019}
S.~C. Surace and J.-P. Pfister, ``Online maximum-likelihood estimation of the
  parameters of partially observed diffusion processes,'' \emph{IEEE
  Transactions on Automatic Control}, vol.~64, no.~7, pp. 2814--2829, Jul.
  2019.

\bibitem{benes1981}
V.~E. Beneš, ``Exact finite-dimensional filters for certain diffusions with
  nonlinear drift,'' \emph{Stochastics}, vol.~5, no. 1-2, pp. 65--92, Jun.
  1981.

\bibitem{lipster2001a}
R.~S. Lipster and A.~N. Shiryaev, \emph{\BIBforeignlanguage{eng}{Statistics of
  {Random} {Processes} {I}: {General} {Theory}}}, 2nd~ed., ser. Stochastic
  {Modelling} and {Applied} {Probability}.\hskip 1em plus 0.5em minus
  0.4em\relax Berlin Heidelberg: Springer, 2001, no.~5.

\bibitem{borkar2022}
V.~S. Borkar, \emph{Stochastic Approximation: A Dynamical Systems Viewpoint},
  2nd~ed.\hskip 1em plus 0.5em minus 0.4em\relax Singapore: Hindustan Book
  Agency (Springer), 2022.

\bibitem{benoist2021}
T.~Benoist, M.~Fraas, Y.~Pautrat, and C.~Pellegrini, ``Invariant measure for
  stochastic {Schrödinger} equations,'' \emph{Annales Henri Poincaré},
  vol.~22, no.~2, pp. 347--374, Feb. 2021.

\bibitem{benoist2023}
T.~Benoist, J.-L. Fatras, and C.~Pellegrini, ``Limit theorems for quantum
  trajectories,'' \emph{Stochastic Processes and their Applications}, vol. 164,
  pp. 288--310, Oct. 2023.

\bibitem{hazan2022}
E.~Hazan, \emph{Introduction to {Online} {Convex} {Optimization}}, 2nd~ed.,
  ser. Adaptive {Computation} and {Machine} {Learning}.\hskip 1em plus 0.5em
  minus 0.4em\relax Cambridge, Massachusetts London, England: The MIT Press,
  2022.

\end{thebibliography}

\end{document}